# A Synthetic Dataset for 5G UAV Attacks Based on Observable Network Parameters

Joseanne Viana †‡, Hamed Farkhari *†, Pedro Sebastião †‡,

Sandra Lagén ¶, Katerina Koutlia ¶, Biljana Bojovic ¶, Rui Dinis §‡

†ISCTE – Instituto Universitário de Lisboa, Av. das Forças Armadas, 1649-026 Lisbon, Portugal

*PDMFC, Rua Fradesso da Silveira, n. 4, Piso 1B, 1300-609, Lisboa, Portugal

‡IT – Instituto de Telecomunicações, Av. Rovisco Pais, 1, Torre Norte, Piso 10, 1049-001 Lisboa, Portugal

§FCT – Universidade Nova de Lisboa, Monte da Caparica, 2829-516 Caparica, Portugal

¶ CTTC - Centre Tecnològic de Telecomunicacions de Catalunya (CERCA)

Emails: {joseanne_cristina_viana, hamed_farkhari, pedro.sebastiao}@iscte-iul.pt, {slagen, kkoutlia, bbojovic}@cttc.es, rdinis@fct.unl.pt

*Abstract*—Synthetic datasets are beneficial for machine learning researchers due to the possibility of experimenting with new strategies and algorithms in the training and testing phases. These datasets can easily include more scenarios that might be costly to research with real data or can complement and, in some cases, replace real data measurements, depending on the quality of the synthetic data. They can also solve the unbalanced data problem, avoid overfitting, and can be used in training while testing can be done with real data. In this paper, we present, to the best of our knowledge, the first synthetic dataset for Unmanned Aerial Vehicle (UAV) attacks in 5G and beyond networks based on the following key observable network parameters that indicate power levels: the Received Signal Strength Indicator (RSSI) and the Signal to Interference-plus-Noise Ratio (SINR). The main objective of this data is to enable deep network development for UAV communication security. Especially, for algorithm development or the analysis of time-series data applied to UAV attack recognition. Our proposed dataset provides insights into network functionality when static or moving UAV attackers target authenticated UAVs in an urban environment. The dataset also considers the presence and absence of authenticated terrestrial users in the network, which may decrease the deep network's ability to identify attacks. Furthermore, the data provides deeper comprehension of the metrics available in the 5G physical and MAC layers for machine learning and statistics research. The dataset will available at link https://archive-beta.ics.uci.edu/

## I. INTRODUCTION

### A. Motivation

Unmanned Aerial Vehicle (UAV) network security is becoming increasingly critical with the rapid expansion in 5G UAV use cases. As the authors in [1] show, all UAV network systems suffer from security flaws that are both technically and commercially demanding for manufacturers to resolve. Common issues are that UAVs can be hacked by disrupting the communication between the UAV and small cell in a jamming attack, and the gps link is also vulnerable to attacks. Most of today's use cases for UAVs apply to logistics and searches such as high-value goods transportation and dynamic object identification, which makes the urban environment suitable for experimenting and testing UAV behavior in the network. Moreover, this is the most common environment for UAV emergency delivery for medical logistics, where the ability to communicate safely and reliably is crucial. The UAVs capability to determine and respond to threats is of the utmost importance.

Deep learning (DL) algorithms are a suitable technique to use for security as they optimize the input parameters to a universal function at scale. Specifically, they offer a broad range of capabilities like excellent learning ability, parallel implementation and in-depth reasoning that makes them ideal for addressing complex problems in temporal modeling [2, 3, 4]. Additionally, machine learning researchers might wish to improve deep networks by decreasing the number of trainable parameters, deep network sizes, and memory requirements, or they may want to speed up training and propose new architectures for better performance. However, deep learning models' accuracy directly depends on the quality and amount of training data available. In some cases, there is no data or the publicly available data is not enough for training a new deep learning architecture especially in the area of 5G UAV communication security. It is possible to generate synthetic data automatically and in endless amounts, unlike actual data, which is costly, time-consuming, and labor-intensive to gather and annotate. Synthetic data can generate balanced datasets that include all plausible real-world data variations for training deep learning models. These models may potentially provide an alternative to data-intensive deep learning methods for UAV attack detection [5] for two reasons: First, user information is kept private which is an important feature. Second, it facilitates direct comparisons between various DL approaches by employing a common data source. For these reasons researchers are looking at the capacity of Convolutional Neural Networks (CNNs) to learn from synthetic data. While dataset availability for training and testing deep learning algorithms is quickly progressing in other UAVs applications, such as UAVs computer vision monitoring [6, 7], we consider that the following obstacles are slowing the progress of UAV network security:

- The lack of up-to-date datasets on attacks based on the 5G network configuration using modern approaches. In the case of UAVs, there are no publicly available datasets.

- In most of the cases, there is a knowledge gap between DL and telecommunication researchers: DL researchers lack understanding of networking, while telecommunication researchers lack knowledge of DL algorithm development.
- The lack of reliable public datasets. Researchers have to develop and use their environments to create datasets individually. Some use network simulators and others use real-world hardware, such as testbeds, but there is no disclosure of the datasets.

These problems prevent direct comparisons between published algorithms and tracking the progress of DL algorithm development for UAVs. Despite our best efforts, we could not locate any datasets dedicated to UAV attacks that use observable network metrics. To address this gap, we intend to provide machine learning researchers with a dataset for UAV attacks, which they may use to improve detection and identification accuracy in future communication systems. ns3 is particularly useful for implementing UAV scenarios due to the proof of reliability in 5G 3GPP reference scenarios, verified in [8], where the authors have calibrated the 5GLENA simulator under 3GPP reference scenarios guaranteeing the resemblance of the results to that of industrial simulators and real networks. Also, the UAV scenarios are implemented according to 3GPP TR 36.777 in [9] and 3GPP TR 38.901 in [10] standards.

*B. Contribution*

This paper aims to disseminate observable network data and engage the DL community in the research on attacks in 5G and 6G UAV networks. In order to do that, we provide:

- A synthetic dataset for UAV attack detection that uses observable parameters in the 5G communication system, such as the Signal to Interference-plus-Noise Ratio (SINR) and the Received Signal Strength Indicator (RSSI) in different network configurations.
- Synthetic data resources for under-supported universities and labs to have machine learning insights and development.
- Contribution to understanding the data collected in the UAV from the network; hence, non-professionals in telecommunications can comprehend the meaning of the control data.

Notably, we provide the raw data so that machine learning researchers can use any pre-processing techniques on the data, such as downsampling or upsampling.

*C. Paper Structure*

The rest of the paper is structured as follows. Section II reviews the 5G-LENA module used to perform the simulations and extract the parameters. Section III details the UAV configuration and the tested scenario. Section IV describes the created synthetic open-source dataset and presents the various extracted metrics for different configuration scenarios. Finally, Section V concludes this paper.

II. THE NS-3 5G-LENA MODULE ns-3 is an open-source GNU GPLv2 licensed software that simulates widely used network and telecommunication protocols in centralized and decentralized scenarios for research purposes. It is maintained and updated by a worldwide community of users [11]. The 5G-LENA is a new module available in the ns-3 network-level simulator that simulates the 5G NR technology, based on 3GPP Release-15 [12].

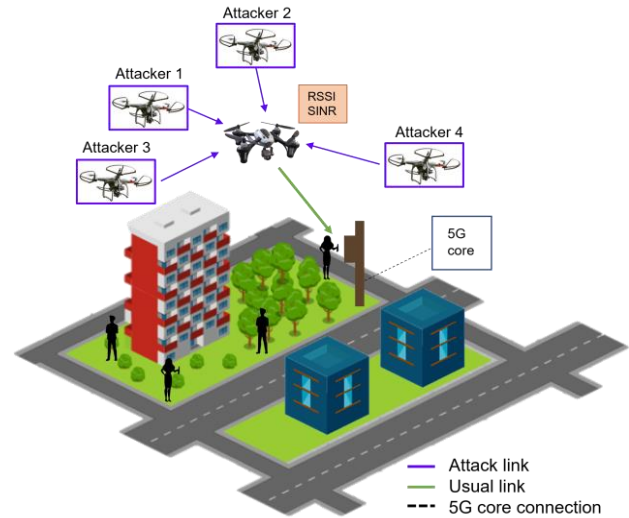

Fig. 1: Simulation environment.

The 5G-LENA module implements a high-fidelity full protocol stack using a non-standalone architecture, i.e., it simulates the 5G radio access network, anchored to a 4G packet core network. It includes the support of various numerologies and flexible frame structures, multiple bandwidth parts (BWPs), time division duplexing (TDD) and frequency division duplexing (FDD) modes, Orthogonal Frequency Division Multiple Access (OFDMA) and Time Division Multiple Access (TDMA) access modes with variable transmission time intervals, dynamic time and frequency domain scheduling, real and ideal beamforming methods, NR-compliant processing timings, and dual-polarized MIMO, among others. It supports operation in the 0.5-100 GHz bands, according to the 3GPP spatial channel model in [10], with the use of 3GPP-compliant uniform planar arrays. The data channels use Low-Density Parity-Check codes (LDPC). The adaptive modulation scheme can generate up to 256 QAM modulation. Hybrid Automatic Repeat reQuest (HARQ) is supported, by using either Incremental Redundancy or Chase Combining methods for receiver combining.

The 5G-LENA strives to imitate reality as closely as possible. For instance, the 5G-LENA simulator has been recently calibrated under 3GPP reference scenarios in [8] guaranteeing results analogous of industrial simulators and real networks. To reduce computational complexity, it applies several basic principles and abstraction models to emulate complex structures, such as the physical layer (which is abstracted by means of a link-to-system mapping [13]), reference signals available in the 5G protocol (such as the Sounding Reference

Signals [14]), and simplified rank adaptation algorithms for MIMO [15].

### III. UAV Configuration Scenarios

Our experiment simulates 1km x 1km area where authenticated UAVs are flying considering fast fading and small fading channels in UAV scenarios added to the 5G-LENA module as in 3GPP [9]. We have created configurations setups based on four urban outdoor scenarios. Namely None speed, Attack speed, User speed, and Both speed. In all configurations, there is a finite number of authenticated UAVs connected to a small cell, over a wireless channel. Also, there is a finite number of attackers and a finite number of terrestrial users. Figure 1 illustrates an instance of a deployment scenario, for 1 authenticated UAV, 4 attackers, and 4 ground users.

Depending on the simulation parameters, UAV attackers might try to degrade the UAV air-to-ground (A2G) link to increase data loss or disrupt the communication using decentralised capabilities and as few resources as possible, while the authenticated UAVs are connected to the small cells and transmit downlink signals. In this case, the attackers use the same propagation models as the authenticated UAVs. Throughout the simulation, the jamming UAVs have the ability to adjust their position.

When the simulation is configured with common terrestrial users, the network link between the users and the small cells follows the 5G traditional infrastructure to exchange downlink data over the internet [10]. The terrestrial users ask for connection to the closest small cell.

From the point of view of the authenticated UAVs, the attackers are in fixed positions or flying in unknown locations inside the coverage area in the air and they can deliberately jam the signal received by the authenticated UAV. In moving attacker scenarios, the attackers' speed is constant always targeting the authenticated UAVs (i.e. they move towards the authenticated UAV). The altitudes of the attackers and the authenticated UAV are all greater than the heights proposed in the standards in order to validate the UAV scenario configuration. The terrestrial users are located in random positions for all scenarios, but they can only move in the User speed and the Both speed scenarios and their new positions update according to [16]. The speed configurations remain the same as in previous scenarios (i.e., attackers speed and users' speed). The RSSI and SINR are collected at the authenticated UAV, in all the configuration scenarios. Both observable parameters are estimated based on the link between the authenticated UAV and the small cell. Table I describes the general parameters of the scenarios in the dataset. The distance in Table I refers to the distance between the small cell and the authenticated UAV. The scenarios also emulate small cells with a height of 10 m and normal traffic.

This dataset verifies end-to-end jamming attacks by checking the RSSI and SINR parameters. The terrestrial users were added to the experiment to increase the scenario complexity, as the authenticated UAV cannot recognize if the interference comes from random users or is an intentional attack. We elaborate the datasets for line-of-sight (LoS) conditions and other datasets associated with UAV scenarios.

### IV. The Dataset

We have created a dataset containing 4800 files divided into half between RSSI and SINR measurements on the receiver side of the authenticated UAVs connected to the network that may or may not have attackers. In what follows we present the tree structures of the dataset: (i) The UAV attacks folder, which is the main folder that contains the whole dataset, (ii) four folders denominated None speed, Attack speed, User speed, and Both speed where all the configurations of the respective scenarios are located, (iii) a total of 600 TABLE I: Network Parameters.

| Scenario Parameters | Value |
|---|---|
| Terrestrial Users | 0,3,5,10,20,30 |
| Authenticated UAVs | 1 |
| Small Cells | 10 |
| Small cell height | 10m |
| Attackers | 0,1,2,3,4 |
| Speeds | 10 km/h |
| modulation scheme | OFDM |
| Small cell power | 4dBm |
| Authenticated UAV power | 2dBm |
| Attackers power | 0,2,5,10,20 dBm |
| Authenticated UAV position | random |
| Attackers position | random |
| Small cells position | random |
| Scenario | UMi |
| Distance | 100,200,500,1000 m |
| Simulation time | 20s |

File: None speed file
```
0.103357,-80.2106,0,0,0,0,0,0,100
0.104357,-80.2106,0,0,0,0,0,0,100
```

File: Attack Speed file
```
0.103357,-81.8423,0,0,1,10,0,0,100
0.104357,-81.8423,0,0,1,10,0,0,100
```

File: User speed file
```
0.103357,-83.0043,0,0,0,0,0,1,100
0.104357,-83.0043,0,0,0,0,0,1,100
```

File: Both Speed file
```
0.103357,-85.4996,0,0,1,10,0,100
0.104357,-85.4996,0,0,1,10,0,100
```

Fig. 2: RSSI.txt head example in every scenario.

configuration folders inside each scenario (see explanation in the previous section) and (iv) the RSSI and SINR time-series data files resulting from the simulation.

The configuration folder's name describes the scenario in which the simulation happened, and the box below shows the structure of the folders names.

attacker_<*attacker number*>_Apower_<*attacker power*>_users_<*terrestrialusers*>_distance_<*distance*>

For example in the folder: attacker_0_Apower_0_users_0_distance_100, there are no attackers, the power of each attacker is set to 0 dBm, there are no terrestrial users, and the distance between the small cell and the Authenticated UAV is 100 meters.

In this other example, attacker_4_Apower_5_users_20_distance_500, the authenticated UAV and the small cell are 500 meters apart. There are four attackers, each with a power of 5 dBm and 20 terrestrial users.

The RSSI.txt and SINR.txt list the following nine comma-separated values:

Variables: *Time*, *<file parameter>*, *Attacker numbers*, *Attacker power*, *Moving attackers*, *Attacker speed*, *User numbers*, *Moving users*, *Distance*

- Time is the instantaneous simulation time when the parameters are measured. The sampling rate is 1ms.
- File parameter is the value of SINR or RSSI (in dBm) collected depending on the file that is being used.
- Attacker numbers is the number of attackers in the simulation.
- Attacker power is the power of each attacker in the simulation.
- Moving attackers is a boolean value to specify if the attackers are moving or not (i.e., 0 is not moving, 1 is moving).
- Attacker Speed defines the speed of each attacker in the

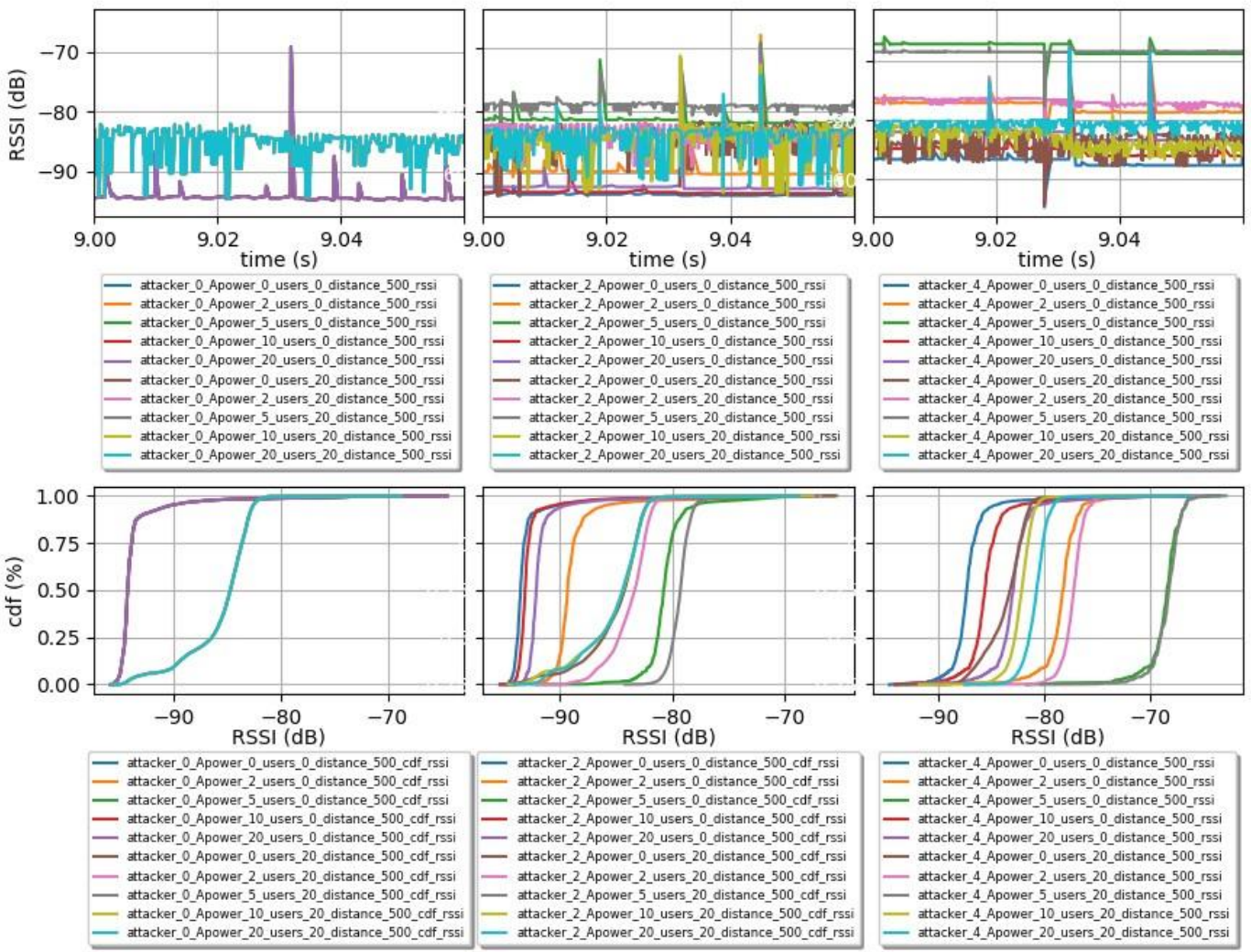

Fig. 3: RSSI in the user speed configuration for different attackers number, attacker powers and users number, for distance of 500 m. simulation when the attacker is moving.

- Users number defines the number of terrestrial users in the simulation.
- Moving users is a boolean value to specify if the users are moving (i.e., 0 is not moving, 1 is moving).
- Distance defines the distance between the small cell and the authenticated UAV.

The labels available in both files are the same because they are related to the same simulation. Fig. 2 illustrates a small piece of the RSSI.txt file available in the dataset configured in the different scenarios.

The usual way to process the variables in the files is reading all of them using dataframes in python. It is possible to use pre-processing techniques, such as sampling and conversion to supervised learning before feeding the data to the deep network.

Fig. 3 summarizes the time domain and the Cumulative Distribution Function (CDF) of the RSSI results in different configurations for the user speed scenario. In the top three figures, we give only a small fraction of the time domain data. As it can be seen from the figures, the interference (RSSI values) increases with the attackers power and with the number of users available in the simulation (charts results are moving of 100 m.

to the right), as expected. The first figure on the left confirms that the attackers power does not affect the overall results when there are no attackers in the scenario. The time domain and CDF results depict the difference between configurations with no users and 20 users only. In the middle chart, when the power is adjusted to five and there are 20 users, the RSSI values increment is higher, while the RSSI has the lowest values when the power is readjusted to 0 and there are no users. The fact that the simulation with the highest power might not present the highest RSSI values in some cases is justified by the fluctuations in the 3GPP channel models (i.e. fading) during each simulation.

Fig. 4 presents the SINR in the time domain and its CDFs in different configurations when the distance between the authenticated UAV and the small cell is 100 m. We present a subset of the time domain values in the top three pictures. Concerning to SINR, the overall CDFs shift to the opposite side of the charts. They move to the left, indicating the decrease in the reception quality of the authenticated UAV as the number of attackers, attackers' power, and users increases. A similar fluctuation of RSSI in the time domain is seen in the SINR. This is due to the stochastic character of the channel model utilized in the simulation. Note that the dataset provides data over time, while we use CDFs here to explain the general concepts behind the observable parameters.

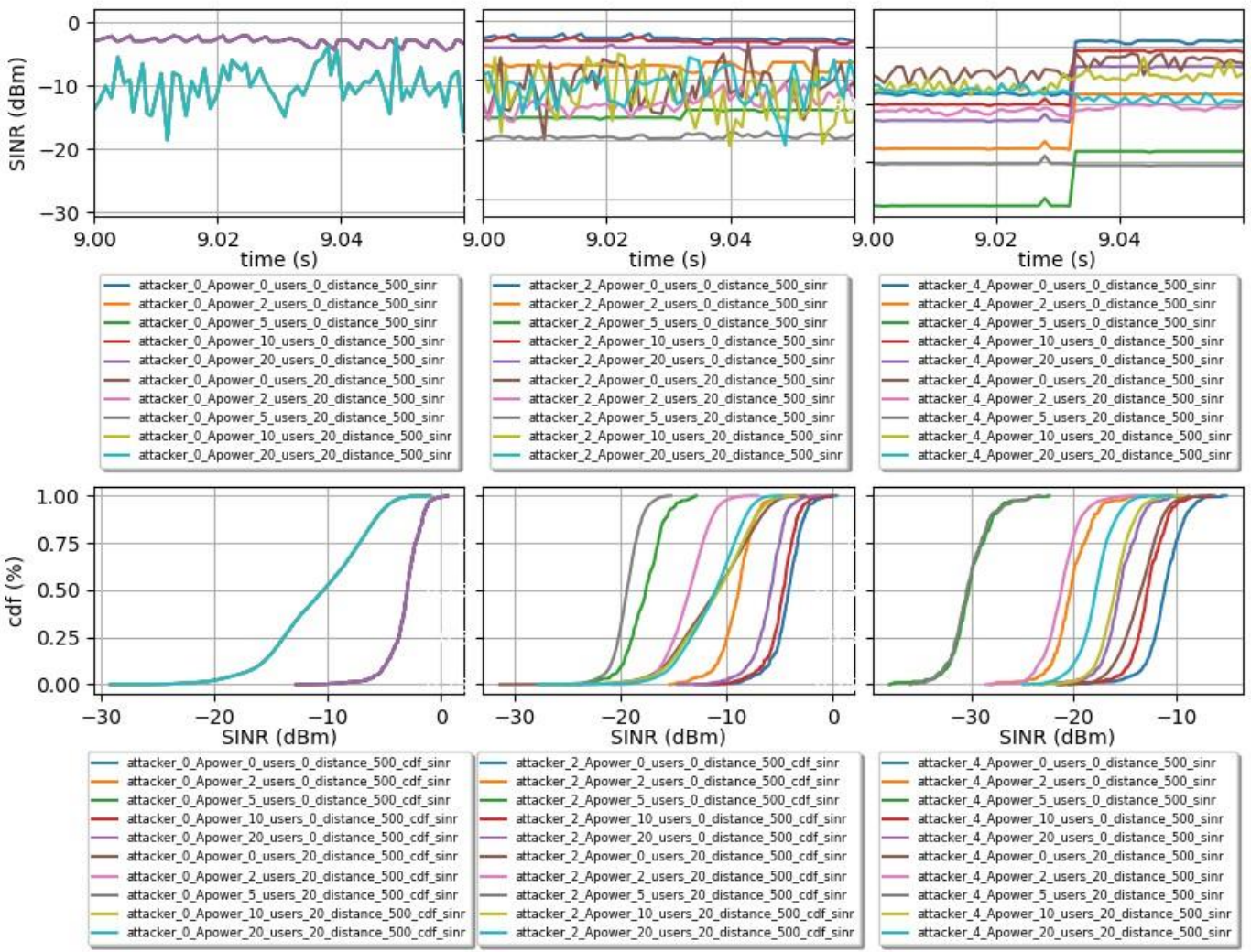

Fig. 4: SINR in the user speed configuration for different attackers number, attacker powers and users number, for distance

*A. Application Example*

The paper in [17] presents a possible application for the dataset where the authors propose a small multiheadedattention-based deep network ( trainable parameters « 1M) for jamming identification that can be embedded in UAVs when the network transmission is configured with Orthogonal Frequency Division Multiplexing - (OFDM) receivers on Clustered Delay Line - (CDL) channels, and the UAVs utilize 3GPP models to estimate network losses. The author discusses concerns regarding robustness and coverage of the deep network when different parameters and analytic contexts are considered, as well as the accuracy of the deep network in such scenarios.

*B. How to use the dataset*

The dataset is composed of files with labels. Usually, machine learning engineers would read the files and convert them to supervised or unsupervised data files learning and apply the learning model to the train and test data depending on the application and proposed analysis. It is possible to use techniques, such as normalization and data reduction in the pre-processing step. The dataset provides the data over time.

## V. CONCLUSION

This paper has provided the first publicly available synthetic dataset for UAV attacks in 5G networks based on two key observable parameters (RSSI and SINR), extracted using the ns-3 5G-LENA simulator. The dataset size is approximately 12GB, compressed in a roughly 800MB size file, and can be accessed through the link https://archive-beta.ics.uci.edu/. The dataset considers attackers and terrestrial users in fixed and moving positions for a wide range of distances between the authenticated UAVs and the small cell. In the future, we plan to extend the set of available UAV scenarios and add more functionalities in the network to increase the generalization, accuracy, and robustness of our dataset compared to real-world data, so that the machine learning researchers can develop and test new algorithms.

## ACKNOWLEDGMENT

This research received funding from the European Union's Horizon 2020 research and innovation programme under the Marie Sklodowska-Curie Project Number 813391 and ANEMONE (PID2021-126431OB-I00) project by the Spanish Government.

## REFERENCES

[1] Muktar Yahuza et al. "Internet of Drones Security and Privacy Issues: Taxonomy and Open Challenges". In: *IEEE Access* 9 (2021), pp. 57243–57270. DOI: 10.1109/ACCESS.2021.3072030.

[2] Bendong Zhao et al. "Convolutional neural networks for time series classification". In: *Journal of Systems Engineering and Electronics* 28.1 (2017), pp. 162–169.

[3] Hassan Ismail Fawaz et al. "Deep learning for time series classification: a review". en. In: *Data Min. Knowl. Discov.* 33.4 (July 2019), pp. 917–963.

[4] Qian Mao, Fei Hu, and Qi Hao. "Deep Learning for Intelligent Wireless Networks: A Comprehensive Survey". In: *IEEE Communications Surveys & Tutorials* 20.4 (2018), pp. 2595–2621. DOI: 10.1109/COMST.2018.2846401.

[5] James Jordon, Alan Wilson, and Mihaela van der Schaar. "Synthetic Data: Opening the data floodgates to enable faster, more directed development of machine learning methods". In: *arXiv e-prints* (2020), arXiv–2012.

[6] Elizabeth Bondi et al. "BIRDSAI: A Dataset for Detection and Tracking in Aerial Thermal Infrared Videos". In: *2020 IEEE Winter Conference on Applications of Computer Vision (WACV)*. 2020, pp. 1736–1745. DOI: 10.1109/WACV45572.2020.9093284.

[7] Qian Gao, Xukun Shen, and Wensheng Niu. "Large-Scale Synthetic Urban Dataset for Aerial Scene Understanding". In: *IEEE Access* 8 (2020), pp. 42131–42140. DOI: 10.1109/ACCESS.2020.2976686.

[8] Katerina Koutlia et al. "Calibration of the 5G-LENA system level simulator in 3GPP reference scenarios". In: *Simulation Modelling Practice and Theory* 119 (2022), p. 102580. ISSN: 1569-190X. DOI: https://doi.org/10.1016/j.simpat.2022.102580. URL: https://www.sciencedirect.com/science/article/pii/S1569190X22000697.

[9] *3GPP - Technical Specification Group Radio Access Network; Study on Enhanced LTE Support for Aerial Vehicles*. URL: https://portal.3gpp.org/desktopmodules/Specifications/SpecificationDetails.aspx?specificationId=3231.

[10] *3GPP - Technical Specification Group Radio Access Network; Study on channel model for frequencies from 0.5 to 100 GHz*. URL: https://portal.3gpp.org/desktopmodules/Specifications/SpecificationDetails.aspx?specificationId=3173.

[11] T. R. Henderson et al. "Network simulations with the ns-3 simulator". In: *SIGCOMM demonstration* 14 (2008).

[12] Natale Patriciello et al. "An E2E simulator for 5G NR networks". In: *Simulation Modelling Practice and Theory* 96 (2019), p. 101933. ISSN: 1569-190X. DOI: https://doi.org/10.1016/j.simpat.2019.101933. URL: https://www.sciencedirect.com/science/article/pii/S1569190X19300589.

[13] S. Lagen et al. "New Radio Physical Layer Abstraction for SystemLevel Simulations of 5G Networks". In: *IEEE Int. Conf. on Commun.* 2020.

[14] Biljana Bojovic, Sandra Lagen, and Lorenza Giupponi. "Realistic Beamforming Design using SRS-based Channel Estimate for ns-3 5GLENA Module". In: *Proceedings of the Workshop on ns-3*. WNS3 2021. Association for Computing Machinery, 2021, pp. 81–87.

[15] Biljana Bojovic, Zoraze Ali, and Sandra Lagen. "ns-3 and 5G-LENA Extensions to Support Dual-Polarized MIMO". In: *Proceedings of the Workshop on ns-3*. WNS3 2022. Association for Computing Machinery, 2022.

[16] L. F. HENDERSON. "The Statistics of Crowd Fluids". In: *Nature* 229.5284 (1971), pp. 381–383. DOI: 10.1038/229381a0. URL: https://doi.org/10.1038%2F229381a0.

[17] Joseanne Viana et al. "A Convolutional Attention Based Deep Learning Solution for 5G UAV Network Attack Recognition over Fading Channels and Interference". In: *2022 IEEE 96th Vehicular Technology Conference* (2022). DOI: 10.48550/ARXIV.2207.10810. URL: https://arxiv.org/pdf/2203.11373.pdf.